\documentclass[11pt]{article}

\usepackage{mathtools, amssymb, fontawesome5, fullpage}
    \allowdisplaybreaks
    \def\A{\mathsf A}
    \def\C{\mathsf C}
    \def\G{\mathsf G}
    \def\T{\mathsf T}
    \def\leq{\leqslant}
    \def\geq{\geqslant}
    \def\BBB{\mathcal B}
    \def\RRR{\mathcal R}
    \def\data{\mathsf{data}}
    \newtheorem{theorem}{Theorem}
    \DeclareMathOperator\Bin{Bin}
    \DeclareMathOperator\Poi{Poi}
    \DeclareMathOperator\Prob{Prob}
    \newtheorem{lemma}[theorem]{Lemma}
    \def\semi{\mathrel{\scalebox{1.2}{$;$}}}
    \newtheorem{definition}[theorem]{Definition}
    \DeclareMathOperator\Capa{Cap}\def\Cap{\Capa}
    \newtheorem{proposition}[theorem]{Proposition}
    \def\fish{{\rlap{\raisebox{.1ex}{\scalebox{.5}{$\,\cdot$}}}\propto}}

\usepackage[rgb, dvipsnames, svgnames]{xcolor}
    \definecolorseries{no-y}{hsb}{last}[hsb]{.2,1,1}[hsb]{1.1,1,1}

\usepackage{tikz-cd}
    \usetikzlibrary{decorations.pathreplacing}
    \tikzset{every picture/.style={line cap=round, line join=round}}

\usepackage{snaptodo}
    \snaptodoset{margin block/.style={font=\tiny}}

\usepackage[colorlinks, allcolors=blue!50!magenta!50!black]{hyperref}

\usepackage{autonum}

\begin{document}

                                 \title
                     {Geno-Weaving: Low-Complexity
                    Capacity-Achieving DNA Storage}
                                    
                                \author
                {Hsin-Po Wang and Venkatesan Guruswami}
                                    
                         \def\day#1\year{\year}
                                    
                               \maketitle
                               \thispagestyle{empty}

\begin{abstract}
    As a possible implementation of data storage using DNA, multiple
    strands of DNA are stored in a liquid container so that, in the
    future, they can be read by an array of DNA readers in parallel.
    These readers will sample the strands with replacement to produce a
    random number of noisy reads for each strand.  An essential
    component of such a data storage system is how to reconstruct data
    out of these unsorted, repetitive, and noisy reads.

    It is known that if a single read can be modeled by a substitution
    channel $W$, then the overall capacity can be expressed by the
    \emph{Poisson-ization} of $W$.  In this paper, we lay down a
    rateless code along each strand to encode its index; we then lay
    down a capacity-achieving block code at the same position across all
    strands to protect data.  That weaves a low-complexity coding scheme
    that achieves DNA's capacity.
\end{abstract}

\footnotetext[0\def\thefootnote{\faDollarSign}]{
    Research supported in part by NSF grant CCF-2210823 and a Simons
    Investigator Award.
}

\footnotetext[0\def\thefootnote{\faChalkboardTeacher}]{
    Preliminary results shared in the Coding Theory and Algorithms for
    DNA-based Data Storage Workshop (a satellite workshop of ISIT) in
    July, 2024 at Athens, Greece and the Information Theory and
    Applications Workshop (ITA) in February, 2024 at San Diego,
    California.
}

\footnotetext[0\def\thefootnote{\faEnvelope[regular]}]{
    Emails: hsinpo@ntu.edu.tw, venkatg@berkeley.edu.
}

\section{Introduction}

    DNA molecules are considered a date storage medium with immense
    potential for their nano-scale size and century-level (some say
    millennial) durability.  To improve its reliability and reduce its
    cost, error-correcting codes are proposed to protect the data we
    store on DNA.

    The biochemical nature of DNA, however, raises nontraditional coding
    challenges.  Unlike magnetic tapes, hard drives, or other sequential
    media, a long strand of DNA is difficult to synthesize, maintain,
    and read.  It needs to be broken into short strands---or it breaks
    by itself\footnote{See \cite{ShV21, BMY22} for interesting results
    on uncontrolled breakage of DNA strands.}---to facilitate
    parallelism in both synthesis and data-retrieval.  For the synthesis
    part, we refer readers to \cite{ALD20} and \cite{HVS23} for the
    latest bioengineering techniques and \cite{LLR20, ALY23, CKL23,
    ElH23} for the latest coding-theoretic works.

    In this paper, we focus on the greatest amount of data that can be
    reliably retrieved and how efficiently this can be done---that is,
    we want low-complexity capacity-achieving codes.  To retrieve data
    from DNA, we put an array of nanopores on a chip.  DNA strands will
    run into pores by chance, and a pore will output a read for each
    strand passing through.  During this process, because all DNA
    strands float in liquid, a pore cannot select or predict which
    strand is coming.  In particular, a pore cannot reject strands that
    were already read.

    Said implementation of DNA data storage is currently the most
    promising one, of which readers can find more details in
    \cite{XYR20, WZB21, YHZ22}.  It also has been studied by many
    coding-theoretic works via the following mathematical model.

\pgfmathdeclarerandomlist{ACGT}{{\A}{\C}{\G}{\T}}
\pgfmathdeclarefunction{wavex}{1}{\pgfmathparse{
    sin(\phasex + \frequx*#1) * 0.1
}}
\pgfmathdeclarefunction{wavey}{1}{\pgfmathparse{
    #1 + sin(\phasey + \frequy*#1) * 0.1
}}
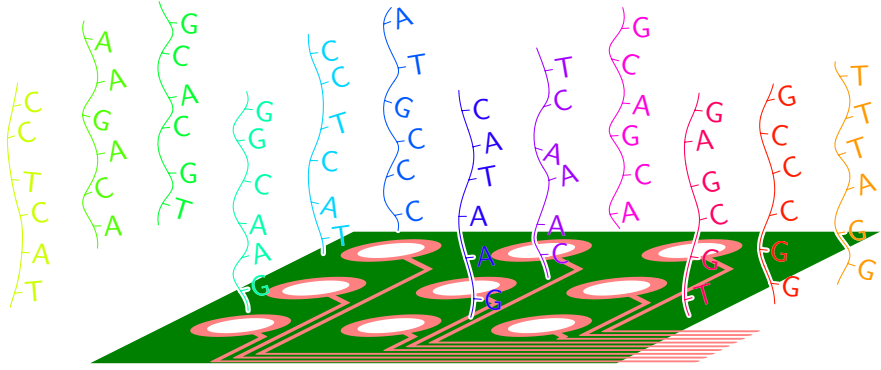
\begin{figure}
    \centering
    \begin{tikzpicture}
        \begin{scope}[transform canvas={cm={1, 0, 0.5, 0.25, (1, 0)}}]
            \fill [green!50!black, even odd rule]
                (0, 0) rectangle (7, 7)
                foreach \x in {1, 3, 5} {
                    foreach \y in {2, 4, 6} {
                        (\x, \y) circle [radius=0.5cm]
                    }
                }
            ;
            \draw [red!50!white, line width=0.2cm]
                foreach \x in {1, 3, 5} {
                    foreach \y in {2, 4, 6} {
                        (\x, \y) circle [radius=0.5cm]
                    }
                }
            ;
            \draw [red!50!white, line width=0.1cm]
                foreach \x in {1, 3, 5} {
                    foreach \y in {2, 4, 6} {
                        (\x, \y) circle [radius=0.5cm]
                        + (-45:0.5) -- (\x+0.4+\y*0.1, \y-0.4-\y*0.1)
                        -- (\x+0.4+\y*0.1, -0.4+\x*0.3+\y*0.1)
                        -- (8, -0.4+\x*0.3+\y*0.1)
                    }
                }
            ;
        \end{scope}
        \path(0, 0);
        \tikzset{y=0.5cm}
        \pgfmathsetseed{8881616}
        \resetcolorseries[11]{no-y}
        \foreach \x in {0, ..., 11}{
            \tikzset{shift={(\x, rnd*4)}}
            \pgfmathsetmacro\phasex{rnd*360}
            \pgfmathsetmacro\frequx{random(50, 200)}
            \pgfmathsetmacro\phasey{rnd*360}
            \pgfmathsetmacro\frequy{random(50, 200)}
            \foreach \y in {1, ..., 6}{
                \pgfmathrandomitem\a{ACGT}
                \draw [white, line width=2pt]
                    plot [domain=-0.5:0.5, samples=10, variable=\t]
                    ({wavex(\y+\t)}, {wavey(\y+\t)})
                    [shift={({wavex(\y)}, {wavey(\y)})}]
                    [rotate=rand*20, transform shape]
                    (0, 0) -- (0.1, 0)
                ;
            }
        }
        \pgfmathsetseed{8881616}
        \resetcolorseries[11]{no-y}
        \foreach \x in {0, ..., 11}{
            \tikzset{shift={(\x, rnd*4)}}
            \pgfmathsetmacro\phasex{rnd*360}
            \pgfmathsetmacro\frequx{random(50, 200)}
            \pgfmathsetmacro\phasey{rnd*360}
            \pgfmathsetmacro\frequy{random(50, 200)}
            \foreach \y in {1, ..., 6}{
                \pgfmathrandomitem\a{ACGT}
                \draw [{no-y!![\x]}]
                    plot [domain=-0.5:0.5, samples=10, variable=\t]
                    ({wavex(\y+\t)}, {wavey(\y+\t)})
                    [shift={({wavex(\y)}, {wavey(\y)})}]
                    [rotate=rand*20, transform shape]
                    (0, 0) -- (0.1, 0) node [right, inner sep=0]
                    {%
                        \pgfsetlinewidth{2pt}\pgfsetcolor{white}%
                        \rlap{$\a$}%
                        \pgfsetcolor{no-y!![\x]}$\a$%
                    }
                ;
            }
        }
    \end{tikzpicture}
    \caption{
        DNA strands float in liquid; nanopores will read them in
        parallel.  See \cite[Fig.~1]{WZB21} for a more accurate picture.
    }                                                   \label{fig:pore}
\end{figure}

\subsection{Mathematical model of DNA coding}

    Let there be $n$ strands
    \begin{gather}
        X^1, X^2, \dotsc , X^n \in \{0, 1\}^\ell
        \\ \text{(sometimes $\{\A, \C, \G, \T\}^\ell$,
        more generally $[q]^\ell$)}
    \end{gather}
    floating in a liquid container; $\ell$ is the length of each strand.
    The pores will produce $m$ reads (we cut the power after collecting
    $m$ reads)
    \[ Y^1, Y^2, \dotsc, Y^m \in \Sigma^\ell \]
    where $\Sigma$ is the output alphabet of nanopore\footnote{ In an
    actual implementation of nanopores, the output symbols are voltages
    because nucleotides affect how charged ions flow through the pore.
    In this paper, however, we model it as an abstract set $\Sigma$.}.
    Each $Y^r$, for $r \in [m]$, is a noisy read of $X^{S(r)} \in [n]$,
    where
    \[ S \in [n]^{[m]}\]
    is a random function sampled uniformly from all functions that map
    the read indices $[m]$ to the strand indices $[n]$.  That is to say,
    for each $X^{S(r)}$ is a strand sampled independently and uniformly
    from $X^1, X^2, \dotsc, X^n$.  Note that $S(1)$ might equal to
    $S(2)$.  The function $S$ captures the \emph{sampling with
    replacement} behavior of the nanopores.

    The reading noise is modeled by a substitution channel
    \begin{gather}
        W\colon \{0, 1\} \to \Sigma
        \\ \text{(Or $\{\A, \C, \G, \T\}$ or $[q]$)}
    \end{gather}
    that acts memorylessly on each of the $\ell$ letters of each of
    the $m$ reads.  That is, if $X^{S(r)}_p$ denotes the $p$th letter of
    the $S(r)$th strand and $Y^r_p$ denotes the $p$th letter of the
    $r$th read, then
    \[ \Prob\bigl( Y^r_p = y \,|\, X^{S(r)}_p = x \bigr) = W(y|x), \]
    for all input letter $x$, all output letters $y \in \Sigma$, all
    read indices $r \in [m]$, and all position indices $p \in [\ell]$.

    Now that the noise model is set, the first problem is with the
    amount of information this system can carry.

\subsection{Understanding the capacity}

    A sequence of works \cite{HSR17, LSW19, LSW20, ShH21, LHS22, WeM22,
    LSW23} aimed to maximize and upper bound the mutual information
    \[ I(X^1, \dotsc, X^n \semi Y^1, \dotsc, Y^m) \]
    to find the capacity of this storage model.  Roughly speaking, the
    capacity can be attributed to these three factors.
    \begin{itemize}
        \item (Factor-W) The single-letter channel model $W$.
        \item (Factor-S) The fact that $S$ samples indices in $[n]$
            \emph{with replacement} means that some $X^s$ will be
            sampled multiple times, some not at all; it's all random.
        \item (Factor-P) Permuting the strands $X^1, \dotsc, X^n$ does
            not change the distribution of reads, i.e., information
            encoding ordering is lost.
    \end{itemize}

    A popular approach (e.g., \cite{HSR17, LSW19a, SRB20, BBY23, WML23})
    to counter (Factor-P) is to prefix $X^s$ by $\log_q N$ letters that
    represent the strand index $s$ in base $q$, for each $s \in [n]$.
    Let us first assume\footnote{ This assumption is not as impractical
    as it might have sounded.  The error rate of DNA is low (around
    $10^{-4}$) to begin with, and the error rate of the first few
    symbols of a strand is even lower.} that the indexing part is
    completely error-free, meaning that we can always recover $S$,
    eliminating the need to address (Factor-P).

    Now that we are left with (Factor-W) and (Factor-S), i.e., $W$ and the sampling
    behavior of $S$, we infer that $S$ will sample $X^1, \dotsc, X^n$
    for a total of $K^1, \dotsc, K^n$ times, respectively, where the
    $K$'s are random variables following the multinomial distribution
    with $m$ balls and $n$ bins.  This motivates the notion of multi-read
    channels, defined below.
    
    \begin{definition}
        For a $k \in \{1, 2, 3, \dotsc\}$, the $k$-read version $W^k$ of
        a channel $W$ is a channel with the same input alphabet as $W$
        (could be $\{0, 1\}$ or $\{\A, \C, \G, \T\}$ or $[q]$) and
        output alphabet $\Sigma^k$.  Its transition probabilities are
        \[
            W^k(y_1, y_2, \dotsc, y_k \,|\, x) \coloneqq
            W(y_1|x) W(y_2|x) \dotsm W(y_k|x)
        \]
        for all inputs $x$ and outputs $y_1, y_2, \dotsc, y_k \in
        \Sigma$.
    \end{definition}

    As both $m$ and $n$ go to infinity with a fixed ratio $\lambda
    \coloneqq m/n$, the multinomial distribution can be approximated by
    the Poisson distribution with intensity $\lambda$.  (See
    \cite[Figure~3a]{OAC18}.) We denote its p.m.f. by $\Poi_\lambda$,
    i.e.,
    \[ \Poi_\lambda(k) \coloneqq \frac{\lambda^k}{k!e^\lambda} \]
    for $k \in \{0, 1, 2, \dotsc\}$.  This leads to the definition of
    \emph{Poisson-ization}.  The capacity of the Poisson-ization,
    $\Cap(W^{\fish\lambda})$, is closely related to the capacity of the
    storage model descried above.

    \begin{definition}
        The Poisson-ization of a channel $W$ is defined as
        \[
            W^{\fish\lambda} \coloneqq
            \sum_{k=0}^\infty \Poi_\lambda(k) W^k.
        \]
        It is a channel with the same input alphabet as $W$ and output
        alphabet $\Sigma^\star \coloneqq \bigcup_{k=0}^\infty \Sigma^k$.
        For each $k \in \{0, 1, 2, \dotsc\}$, the output of
        $W^{\fish\lambda}$ is a $k$-tuple, and will be generated by
        $W^k$, with probability $\Poi_\lambda(k)$.
    \end{definition}

    \begin{definition}
        The capacity of a channel $W$ is denoted as $\Cap(W)$.
    \end{definition}

    \begin{theorem}                                    \label{thm:withS}
        Suppose\footnote{ This is equivalent to assuming that each $X^s$
        spends $\log_q n$ letters of their length $\ell + \log_q n$ to
        encode the strand index $s \in [n]$, and the indexing part is
        error-free.} that a genie reveals $S$.  Suppose that, as $\ell$,
        $m$, and $n$ increase, $\lambda \coloneqq m/n$ remains a
        constant.  Then
        \[
            I(X^1, \dotsc, X^n \semi Y^1, \dotsc, Y^m, S) <
            (1 + o(1)) n \ell \Cap(W^{\fish\lambda}).   \label{eq:withS}
        \]
        That is, the capacity of DNA coding is determined by the
        Poisson-ization of the channel $W$ that models single-letter
        errors.  (Note: We let $I$ and $\Cap$ assume the same log-base.
        Readers can choose their favorite base.)
    \end{theorem}

    Theorem~\ref{thm:withS} can be proved by a standard Shannon-type
    argument, which will be reproduced in Section~\ref{sec:withS} for
    readers' convenience.  Some works use
    \[
        \sum_{k=0}^\infty \Poi_\lambda(k) \Cap(W^k)    \label{eq:poicap}
    \]
    to upper bound the $\Cap(W^{\fish\lambda})$ on the right-hand side
    of \eqref{eq:withS}.  We remark that the equality does not hold for
    when $W, W^2, W^3, \dotsc$ do not share the same capacity-achieving
    input distributions.  In particular, if $W$ is not symmetric and
    the uniform distribution is not capacity-achieving,
    \eqref{eq:poicap} is probably strictly greater than
    $\Cap(W^{\fish\lambda})$.

    A contribution of ours is to propose a low-complexity coding scheme
    that saturates Theorem~\ref{thm:withS}.

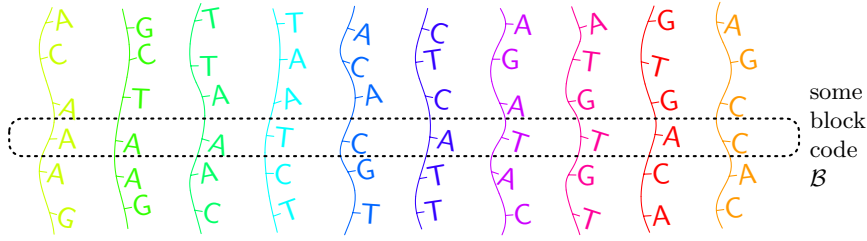
\begin{figure}
    \centering
    \begin{tikzpicture}[y=0.5cm]
        \pgfmathsetseed{8882424}
        \resetcolorseries[9]{no-y}
        \foreach \x in {0, ..., 9}{
            \tikzset{shift={(\x, 0)}}
            \pgfmathsetmacro\phasex{rnd*360}
            \pgfmathsetmacro\frequx{random(50, 200)}
            \pgfmathsetmacro\phasey{rnd*360}
            \pgfmathsetmacro\frequy{random(50, 200)}
            \foreach \y in {1, ..., 6}{
                \pgfmathrandomitem\a{ACGT}
                \draw [{no-y!![\x]}]
                    plot [domain=-0.5:0.5, samples=10, variable=\t]
                    ({wavex(\y+\t)}, {wavey(\y+\t)})
                    [shift={({wavex(\y)}, {wavey(\y)})}]
                    [rotate=rand*20, transform shape]
                    (0, 0) -- (0.1, 0)
                    node [right, inner sep=0] {$\a$}
                ;
            }
        }
        \draw [dotted, line width=0.8pt, rounded corners=5pt]
            (-0.5, 3.5) rectangle (10, 2.5)
            (10, 3) node [right, align=left, font=\footnotesize]
            {some \\ block \\ code \\ $\BBB$}
        ;
    \end{tikzpicture}
    \caption{
        To prove Theorem~\ref{thm:useC}, we will apply a block code at
        the same position across all strands to protect data.
    }                                                   \label{fig:useC}
\end{figure}

    \begin{theorem}[presented in ITA 2024]              \label{thm:useC}
        The capacity bound given in Theorem~\ref{thm:withS} can be
        achieved by, for each $p \in [\ell]$, applying a block code from
        a capacity-achieving family at the same position $p$ across all
        strands, as Figure~\ref{fig:useC} illustrates.
    \end{theorem}

    Theorem~\ref{thm:useC} will be proved in Section~\ref{sec:useC}.
    Theorem~\ref{thm:useC} greatly simplifies the problem of DNA coding
    (when somehow we know $S$) to the point that it suffices to find
    error-correcting codes achieving the capacity of $W^{\fish\lambda}$.
    We know there are codes achieving the capacity of any discrete
    memoryless channel (DMC); for instance, polar codes.

    Next we bring (Factor-P) back, but not entirely.  Suppose that the genie
    does not tell us $S$ per se, but tells us the partition\footnote{
    Note that $P \subset 2^{[m]}$ is also a random variable as it
    depends on $S \in [n]^{[m]}$, another random variable.}
    \[
        P \coloneqq
        \{S^{-1}(1), S^{-1}(2), \dotsc, S^{-1}(n)\}
        \subset 2^{[m]}.
    \]
    In other words, the genie performs \emph{clustering} (\cite{SYL22,
    QYW22}) for us, but $P$ as a set lacks ordering---we know
    $S^{-1}(1)$ is a preimage, but it is our job to figure out if it is
    the preimage of $1$ or $2$.  The capacity pays a price for that.

    \begin{theorem}[{\cite[Theorem~3]{ShH21}}]         \label{thm:withP}
        Let $W$ be any symmetric DMC channel.  Suppose that a genie
        reveals $P$.  Suppose that, as $\ell$, $m$, and $n$ increase,
        $\ell / \log n$ remains a constant that is sufficiently large
        and $\lambda \coloneqq m/n$ also remains a constant.  Then 
        \begin{align}
            I(X^1, \dotsc, X^n \semi Y^1, \dotsc, Y^m, P)
            & < (1 + o(1)) n \ell \Cap(W^{\fish\lambda})
            \\ & \mkern3mu - (1 - \Poi_\lambda(0)- o(1)) n \log n.
                                                        \label{eq:withP}
        \end{align}
    \end{theorem}

    That is, the capacity of DNA coding is penalized by the confusion
    caused by shuffling the strands indices $s$, and the logarithm of
    the number of ways to shuffle is $(1 - \Poi_\lambda(0) - o(1)) n
    \log n$.
    
    Theorem~\ref{thm:withP} turns out to be highly nontrivial, costing
    several papers to clarify.  Here is a brief history.  In
    \cite[Theorem~1]{HSR17}, the upper bound is proved for the case when  $W$ is
    noiseless.  In \cite[Theorem~1]{LSW19}, $W$ is a binary symmetric
    channel (BSC).  The channel is then generalized to symmetric DMC
    channels \cite[Theorem~3]{ShH21} and asymmetric DMC channel
    \cite[Theorem~10]{WeM22}.  More recently, \cite[Theorem~4]{LSW23}
    replaces Poisson with general distributions.  In
    Section~\ref{sec:withP}, we will make several comments regarding our
    understanding of Theorem~\ref{thm:withP}.

    In the opposite direction, many works have been trying to achieve
    the upper bound.  For instance, \cite[Theorem~1]{HSR17},
    \cite[Theorem~3]{ShH21}, and \cite[Theorem~4]{LSW23} are ``if and
    only if'' statements; \cite[Theorem~1]{LSW20} achieves the bound of
    \cite[Theorem~1]{LSW19}; \cite[Theorem~1]{LHS22} achieves the
    capacity over binary erasure channels (BECs) with linear codes;
    \cite[Theorem~5]{WeM22} approaches \cite[Theorem~10]{WeM22}.

    This paper's main contribution is to upgrade the low-complexity
    scheme of Theorem~\ref{thm:useC}, so it can now encode ordering
    information.

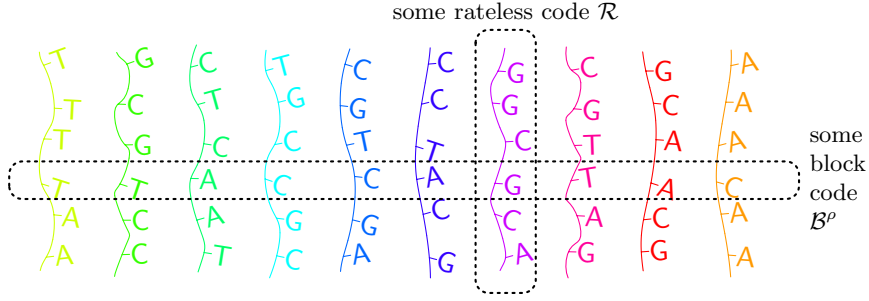
\begin{figure}
    \centering
    \begin{tikzpicture}[y=0.5cm]
        \pgfmathsetseed{8883232}
        \resetcolorseries[9]{no-y}
        \foreach \x in {0, ..., 9}{
            \tikzset{shift={(\x, 0)}}
            \pgfmathsetmacro\phasex{rnd*360}
            \pgfmathsetmacro\frequx{random(50, 200)}
            \pgfmathsetmacro\phasey{rnd*360}
            \pgfmathsetmacro\frequy{\frequx*(1.6 + rnd*0.2)}
            \foreach \y in {1, ..., 6}{
                \pgfmathrandomitem\a{ACGT}
                \draw [{no-y!![\x]}]
                    plot [domain=-0.5:0.5, samples=10, variable=\t]
                    ({wavex(\y+\t)}, {wavey(\y+\t)})
                    [shift={({wavex(\y)}, {wavey(\y)})}]
                    [rotate=rand*20, transform shape]
                    (0, 0) -- (0.1, 0)
                    node [right, inner sep=0] {$\a$}
                ;
            }
        }
        \draw [dotted, line width=0.8pt, rounded corners=5pt]
            (5.7, 7) rectangle (6.5, 0)
            (6.1, 7) node [above, font=\footnotesize]
            {some rateless code $\RRR$}
            (-0.5, 2.5) rectangle (10, 3.5)
            (10, 3) node [right, align=left, font=\footnotesize]
            {some \\ block \\ code \\ $\BBB^\rho$}
        ;
    \end{tikzpicture}
    \caption{
        To prove Theorem~\ref{thm:useBR}, we will deploy a rateless code
        on each strand to encode its index and a block code at each
        position to protect data.  The code rate of the latter will
        depend on $p$.  The letter at any intersection will be the
        mod-$4$ sum of the rateless code symbol and the block code
        symbol.
    }                                                  \label{fig:useBR}
\end{figure}

    \begin{theorem}[presented in ISIT 2024's satellite workshop]
                                                       \label{thm:useBR}
        Let $W$ be a binary memoryless symmetric (BMS)
        channel.\footnote{ For readers not familiar with BMS channels,
        think of BSCs.}\footnote{ For experienced readers interested in
        the most general channels our techniques apply to, think of DMCs
        that are additively symmetric; that is, for any input $a$, there
        is a permutation $\pi_a$ acting on the output alphabet such that
        $W(y|x) = W(\pi_a(y) \,|\, a + x)$.} Let $\ell / \log n$
        remain a constant $> 1 / \Cap(W)$.  The capacity bound in
        Theorem~\ref{thm:withP} can be achieved by combining techniques
        (tech-R) and (tech-B) below, as Figure~\ref{fig:useBR}
        illustrates.  (Tech-R) For each strand, apply a rateless code to
        encode its index.  (Tech-B) Apply a block code from a
        capacity-achieving family at the same position across all
        strands.
    \end{theorem}

    Note that rateless codes have been used in the context of DNA coding
    \cite{ErZ17, AVA19, HeC23}, but never in the same way as we do here.
    Two-dimensional constructions are also very popular among
    implementations and experiments \cite{GHP15, BGH16, YGM17,
    OAC18}---why would it be otherwise?  However, these works are based
    on concatenation of inner and outer codes.  Our decoder, on the
    other hand, follows a zigzag pattern that alternates between
    horizontal and vertical codes.

    The organization of this paper is that Sections \ref{sec:withS},
    \ref{sec:useC}, \ref{sec:withP}, and \ref{sec:useBR} will address
    Theorems \ref{thm:withS}, \ref{thm:useC}, \ref{thm:withP}, and
    \ref{thm:useBR}, respectively.

\section{Capacity Bound of Sampling}                   \label{sec:withS}

    In this section, we want to upper bound
    \[
        I(X^1, \dotsc, X^n \semi Y^1, \dotsc, Y^m, S)
                                                        \label{eq:Ione}
    \]
    to prove Theorem~\ref{thm:withS}.  To begin, we want to match each
    read $Y^r$ to its origin $X^s$ using $S$ as hints.

\subsection{Decomposition of the mutual information}

    For any strand index $s \in [n]$, we let $Y^{S^{-1}(s)} \coloneqq
    (Y^r : r \in S^{-1}(s))$ be the reads related to the $s$th strand
    $X^s$.  Its data type is $(\Sigma^\ell)^{|S^{-1}(s)|}$, a subset of
    in $(\Sigma^\ell)^\star$.  We know that $W$ generated
    $Y^{S^{-1}(1)}$ using $X^1$ without referring to $X^2, \dotsc, X^n$.
    The same can be said to $Y^{S^{-1}(2)}, \dotsc, Y^{S^{-1}(n)}$.  We
    infer that the mutual information \eqref{eq:Ione} is less than or
    equal to
    \[
        I(X^1 {;} Y^{S^{-1}(1)}) + \dotsb +
        I(X^n {;} Y^{S^{-1}(n)}).                     \label{eq:Isum}
    \]

    We can further decompose \eqref{eq:Isum} using position indices: For
    each $p \in [\ell]$, define $Y^{S^{-1}(s)}_p \coloneqq (Y^r_p : r
    \in S^{-1}(s))$.  Its datatype is $\Sigma^{|S^{-1}(s)|}$, a subset
    of $\Sigma^\star$.  We infer that \eqref{eq:Ione} and
    \eqref{eq:Isum} are less than or equal to
    \[ \sum_{s=1}^n \sum_{p=1}^\ell I(X^s_p {;} Y^{S^{-1}(s)}_p). \]
    By symmetry, it suffices to upper bound the single-letter mutual
    information $I(X^1_1 {;} Y^{S^{-1}(1)}_1)$ by, hopefully, the
    capacity $\Cap(W^{\fish\lambda})$.
    
    Clearly such a mutual information is upper bounded by the capacity
    of the channel that outputs $Y^{S^{-1}(1)}_1$ when given the input
    $X^1_1$.  But what exactly is this channel?  It is a channel that
    outputs a $K$-tuple by feeding the input into $K$ i.i.d.\ copies of
    $W$, where $K$ is the number of reads.  As a random variable, $K$
    follows a binomial distribution with $m$ trials and success rate
    $1/n$, which we denote by $\Bin_{m/n}$.  Knowing these, we can write
    \[ I(X^1_1 {;} Y^{S^{-1}(1)}_1) \leq \Cap(W^{m/n}), \]
    where $W^{m/n}$ is the ``binomial-ization'' of $W$
    \[ W^{m/n} \coloneqq \sum_{k=0}^m \Bin_{m/n}(k) W^k. \]
    We can now conclude the finitism version of
    Theorem~\ref{thm:withS}.

    \begin{proposition}                                  \label{pro:bin}
        Fix $\ell$, $m$, and $n$.  Let $K \sim \Bin_{m/n}$.  Then
        \[
            I(X^1, \dotsc, X^n \semi Y^1, \dotsc, Y^m, S)
            \leq n \ell \Cap(W^{m/n}).
        \]        
    \end{proposition}

    It remains to relate $\Cap(W^{m/n})$ to $\Cap(W^{\fish\lambda})$,
    and we can finish Theorem~\ref{thm:withS}.  To be more precise, if
    we can show $\Cap(W^{m/n}) \to \Cap(W^{\fish\lambda})$ as $m, n \to
    \infty$ (while $\lambda \coloneqq m/n$ stays constant), then we are
    done.  We are quite confident about its truthfulness because we
    know $\Bin_{m/n} \to \Poi_\lambda$ and hence $W^{m/n}$ ``$\to$''
    $W^{\fish\lambda}$.  The later limit is quoted because we have not
    specified a topology on the space of channels, let along checking if
    $\Cap$ is continuous w.r.t.\ this topology.  As it turns out, the
    technical details are neither pleasing nor inspiring.  We leave them
    in the next subsection and advise readers to save their attention
    for Sections \ref{sec:useC} and \ref{sec:useBR}.

\subsection{Technical details regarding Poisson limit theorem}
                                                        \label{sec:PLT}

    In this subsection, we want to show that $\Bin_{m/n} \to
    \Poi_\lambda$ implies $W^{m/n}$ ``$\to$'' $W^{\fish\lambda}$
    and $\Cap(W^{m/n}) \to \Cap(W^{\fish\lambda})$.  For that,
    we invoke the following version of Poisson limit theorem.

    \begin{lemma}                                        \label{lem:PLT}
        Let $\Bin_{m/n}$ be the p.m.f.\ of the binomial distribution
        with $m$ trials and success rate $1/n$.  Let $\Poi_\lambda$ be
        the p.m.f.\ the Poisson distribution with intensity $\lambda$.
        If $\lambda \coloneqq m/n$ stays constant while $m, n \to
        \infty$, then $\Bin_{m/n}(k) \to \Poi_\lambda(k)$ for each $k
        \in \{0, 1, 2, \dotsc\}$.  (Note: the speed of convergence might
        depend on $k$.)
    \end{lemma}

    Lemma~\ref{lem:PLT} implies that there is a way to couple random
    variables $K \sim \Bin_{m/n}$ and $K^\fish \sim \Poi_\lambda$ such
    that $\Prob(K \neq K^\fish) \to 0$ as $m, n \to \infty$.

    Denote $\Prob(K \neq K^\fish)$ by $\kappa$.  We see that, with
    probability $1 - \kappa$, both $W^{m/n}$ and $W^{\fish\lambda}$
    output tuples whose length is $K = K^\fish$.  We couple $W^{m/n}$
    and $W^{\fish\lambda}$ further so that even the content of the
    tuples coincide.  That is to say, with probability $1 - \kappa$, one
    cannot feel any difference between $W^{m/n}$ and $W^{\fish\lambda}$.

    With probability $\kappa$, however, the tuples differ in length and
    content, and we might learn more information from the longer tuple.
    Worst case scenario, we will learn everything from $W^{m/n}$
    but nothing from $W^{\fish\lambda}$, or everything from
    $W^{\fish\lambda}$ but nothing from $W^{m/n}$.  We can therefore
    bound $W^{m/n}$ by
    \[
        \Cap(W^{\fish\lambda}) - \kappa \log q
        \leq \Cap(W^{m/n})
        \leq \Cap(W^{\fish\lambda}) + \kappa \log q,
    \]
    where $q$ is the size of the input alphabet of $W$.  We know $\kappa
    \to 0$ as $m, n \to \infty$.  Hence, $\Cap(W^{m/n})$ is sandwiched
    and converges $\to \Cap(W^{\fish\lambda})$.  This together with
    Proposition~\ref{pro:bin} completes the proof of
    Theorem~\ref{thm:withS}.

\section{DNA Coding by Capacity-Achieving Codes}        \label{sec:useC}

    In this section, we use codes that achieve the capacity of the
    Poisson-ization $W^{\fish\lambda}$ to saturate
    Theorem~\ref{thm:withS} and prove Theorem~\ref{thm:useC}.  To be
    more precise, we will deploy a block code $\BBB \subset [q]^n$
    and synthesize DNA strands such that $(X^1_p, X^2_p, \dotsc, X^n_p)
    \in \BBB$ for all positions $p \in [\ell]$.  Here, $X^s_p$ is the
    $p$th letter of the $s$th strand $X^s$.

\subsection{Code selection (technical and skippable)}

    Our plan is to choose a block code $\BBB$ for each length $n$ whose
    code rate approaches the capacity of $W^{\fish\lambda}$ as $n \to
    \infty$.  The plan bumps into a technicality that, at any finite
    $n$, the actual channel in charge is never $W^{\fish\lambda}$ but
    the binomial-ization $W^{m/n}$, as was discussed in
    Section~\ref{sec:PLT}.  Neither can we choose capacity-achieving
    codes for $W^{m/n}$ because the channel itself changes as $n$
    increase.  Off-the-shelf codes usually do not provide capacity
    guarantees in this manner.

    One technical trick that bypasses said technicality is the
    \emph{diagonal argument}.  First, let $Z$ be the zero-capacity
    channel that outputs garbage.  Choose a family of codes that achieve
    the capacity of the compound channel
    \[
        \frac12 W^{\fish\lambda} + \frac12 Z.        \label{eq:compound}
    \]
    For sufficiently large $n$, we can couple $W^{m/n}$ and
    \eqref{eq:compound} such that either they output the same tuple or
    the latter outputs garbage.  Coding-wise, this means that any code
    designed for \eqref{eq:compound} will work equally well, if not
    better, over $W^{m/n}$.

    Next we halve the weight of $Z$ and find a code family that achieves
    the capacity of
    \[ \frac34 W^{\fish\lambda} + \frac14 Z, \]
    Because this channel has a higher capacity than the compound channel
    \eqref{eq:compound} does, there will be a block length $n'$ such
    that, after $n'$, codes in the second family has a higher rate
    than those in the first family.  By the same coupling argument,
    there will also be a block length $n''$ such that codes after $n''$
    perform equally well, if not better, on $W^{m/n}$.  We can then
    switch to using codes from the second family for $n$ greater than
    the maximum of $n'$ and $n''$.

    The diagonal part of this argument is that being able to switch from
    the $Z/2$-family to the $Z/4$-family, we anticipate switching to
    the $Z/8$-family, then the $Z/16$-family, and so on.  Eventually, we
    will obtain a diagonal family that achieves the capacity of
    $W^{\fish\lambda}$ while the actual channels the codes apply to are
    $W^{m/n}$.
    
\subsection{Encoding and code rate}

    Now that we have a family of good codes parametrized by block
    lengths, fix an $n$ and let $\BBB \subset [q]^n$ be the code
    of length $n$ and rate $\rho$.  We know $\rho \to
    \Cap(W^{\fish\lambda})$ as $n \to \infty$ by the previous
    subsection.  Now, for each position $p \in [\ell]$, we will encode
    $\rho n$ amount of data by choosing a codeword
    \[
        (X^1_p, X^2_p, \dotsc, X^n_p) \in \BBB.      \label{eq:codeword}
    \]
    In total, the $\ell$ positions of the $n$ strands carry $\rho n
    \ell$ amount of data.  This clearly is $(1 \pm o(1))$ within the
    right-hand side of \eqref{eq:withS} as $n \to \infty$.

\subsection{Decoding and error probability}

    Recall that $Y^{S^{-1}(s)}_p$ is defined to be the tuple
    $Y^{S^{-1}(s)}_p \coloneqq (Y^r_p : r \in S^{-1}(s)) \in
    \Sigma^{|S^{-1}(s)|} \subset \Sigma^\star$.  For each position
    $p \in [\ell]$, we observe the corrupted word
    \[
        (Y^{S^{-1}(1)}_p,
        Y^{S^{-1}(2)}_p, \dotsc,
        Y^{S^{-1}(n)}_p)
        \in (\Sigma^\star)^n                          \label{eq:corrupt}
    \]
    and use the decoder of $\BBB$ to recover \eqref{eq:codeword}.  If
    with probability $\varepsilon$ we can recover \eqref{eq:codeword}
    from \eqref{eq:corrupt}, then a union bound over all $p \in [\ell]$
    implies that we can recover $X^1, \dotsc, X^n$ from $Y^1, \dotsc,
    Y^r, S$ with error probability $\ell\varepsilon$.

    Generally speaking, good codes are expected to have error
    probabilities $\varepsilon \approx \exp(-\Omega(n))$ (for
    random-like codes) or $\varepsilon \approx \exp(-\sqrt n)$ (for
    polar codes).  The parameter region we are interested in is when
    $\ell \propto \log n$.  We therefore can say that $\ell\varepsilon$
    should be very small.
    
    However, it is unclear if the asymptote $\exp(-\Omega(n))$ can be
    achieved given that the codes were proved existing via a diagonal
    argument---there is no guarantee that the infimum of the
    $\Omega$-constants across different families is positive.  We leave
    that to future works and patch this loophole with an outer code that
    treats one strand as one symbol.  An outer code can correct
    $\varepsilon$ errors if we sacrifice $2\varepsilon$ of the code
    rate.  The diagonal argument does imply that the error probability
    $\varepsilon$ will vanish, which implies that the sacrifice in code
    rate will vanish as well.  That completes the proof of
    Theorem~\ref{thm:useC}.

\section{Capacity Bound of Shuffling}                  \label{sec:withP}

    In this section, we leave some comments on the penalty term in
    Theorem~\ref{thm:withP}
    \[
        \log \frac{n!}{(\Poi_\lambda(0)n)!}
        \approx (1 - \Poi_\lambda(0)) n \log n         \label{eq:penalty}
    \]
    introduced by forgetting $S$.  The first and the obvious comment is
    that we lose $\log(n!)$ amount of information when we cannot recover
    the ordering on $X^1, \dotsc, X^n$.  This explains the dominant term
    $n \log n$ in \eqref{eq:penalty}.

    To explain the $\Poi_\lambda(0)$ term, however, require more
    dedication.  By the Poisson limit theorem, there will be about
    $\Poi_\lambda(0)n$ strands that does not show up in any read.  We
    would not gain any information from them, nor should we worry about
    their ordering.  We therefore penalize the penalty term by putting
    $(\Poi_\lambda(0)n)!$ in the denominator.

    The third and the last comment is that we had assumed there are $n!$
    orderings, which implicitly assumes that $X^s$ are all unique.  This
    may not be the case, as it is not a priori clear that making $X^1 =
    X^2$ will decrease (because $X^2$ does not carry new information) or
    increase (because repetition makes it easier to decode $X^1$) the
    mutual information.  To add more complicacy, notice that when $X^1
    \neq X^2$ but the Hamming distance in between is small, $X^2$ does
    not carry too much information on top of $X^1$ and the ordering
    between them does matter but not so much.

    The work \cite{WeM22} suggests that the upper bound is closely
    related to the \emph{excess capacity} $\Cap(W^{2k}) - \Cap(W^k)$,
    i.e., how much we gain if the reads of $X^2$ are treated as extra
    reads of $X^1$, and its relation to $\ell / \log n$.  A hand-waving
    explanation is that we are choosing between using $X^2$ as a backup
    of $X^1$ (hence $\Cap(W^{2k})$, instead of $\Cap(W^k)$, applies) or
    spending $\log(n) / \Cap(W^k)$ letters to encode a unique index on
    $X^2$ (hence we have $\ell - \log(n) / \Cap(W^k)$ letters left for
    data).

\section{Geno-Weaving with Rateless and Block Codes}   \label{sec:useBR}

    In this section, we want to combine rateless codes with
    capacity-achieving block codes to saturate Theorem~\ref{thm:withP}
    and prove Theorem~\ref{thm:useBR}.  There will be a rateless code
    $\RRR$ and a family of block codes $\BBB^\rho$ parametrized by code
    rates $\rho \in [0, 1]$.  For each strand index $s \in [n]$ and each
    position index $p \in [\ell]$, we will synthesize DNA letter
    \[ X^s_p \coloneqq \RRR(s)_p + \BBB^{\rho(p)}(\data^p)_s. \]
    Here, $\RRR(s)_p$ is the $p$th letter of the rateless stream
    generated by seed $s$, and $\BBB^{\rho(p)}(\data^p)_s$ is the $s$th
    letter of the codeword that encodes the $p$th chunk of our data,
    $\data^p$, whose size is $\rho(p) n$.

    Our attack on Theorem~\ref{fig:useBR} begins with a straw-man proof
    that relies on quite a few unrealistic assumptions.  The goal here
    is to show the core idea without having to worry about deltas and
    epsilons.

\subsection{A straw man with some unrealistic assumptions}

    In this and the upcoming two subsections the following are taken as
    granted.
    \begin{itemize}
        \item The rateless code $\RRR$ is so good that a seed in $[n]$
            can be decoded whenever we have collected $\log(n)/\Cap(W)$
            outputs of $W$, for all BMS channels $W$.
        \item The block codes $\BBB^\rho$ are so good that they can be
            decoded whenever we have collected $\rho n$ outputs of $W$.
        \item We also assume that the binomial distribution that
            controls the read multiplicities is
            \[
                \Bin(0) = 1/4, \qquad
                \Bin(1) = 1/4, \qquad
                \Bin(2) = 1/4, \qquad
                \Bin(3) = 1/4,
            \]
            i.e., the probability of a strand being read $0$, $1$, $2$,
            or $3$ times is $1/4$ each.
        \item Not only that, we will assume that exactly $\Bin(k)n$
            strands will be read $k$ times for $k \in \{0, 1, 2, 3\}$.
    \end{itemize}

\subsection{Encoding and code rates of the straw man}
                                                   \label{sec:strawrate}

    At the beginning of the strands, we use null rate
    \[
        \rho(p) \coloneqq 0
        \qquad\text{ for }\qquad
        p \in \Bigl[
            1,
            \frac{\log_2(n)}{\Cap(W^3)}
        \Bigr].
    \]
    In other words, there should be no data in the first few positions
    until we can decode any indices from the strands that are sampled
    $3$ times.  Let's call these \emph{triple-read strands}.

    We then raise the code rate to
    \[
        \rho(p) \coloneqq \frac{\Cap(W^3)}{4}
        \qquad\text{ for }\qquad
        p \in \Bigl[
            \frac{\log n}{\Cap(W^3)} + 1,
            \frac{\log n}{\Cap(W^2)}
        \Bigr].
    \]
    That is, after we can decode the indices of triple-read strands, and
    before we can decode \emph{double-read strands} (those that are
    sampled $2$ times), we put in a proper amount of data that can be
    decoded from the triple-read strands.

    Following this pattern, we let
    \[
        \rho(p) \coloneqq \frac{\Cap(W^3) + \Cap(W^2)}{4}
        \quad\text{ for }\quad
        p \in \Bigl[
            \frac{\log n}{\Cap(W^2)} + 1,
            \frac{\log n}{\Cap(W^1)}
        \Bigr].
    \]
    That is, after we can decode the indices of triple-read and
    double-read strands we can extract data from them.  Finally,
    \[
        \rho(p) \coloneqq \frac{\Cap(W^3) + \Cap(W^2) + \Cap(W^1)}{4}
        \quad\text{ for }\quad
        p \in \Bigl[
            \frac{\log n}{\Cap(W^1)} + 1,
            \ell
        \Bigr].
    \]
    That is, we go for full capacity after learning the indices of
    \emph{single-read strands} (those that are sampled $1$ time).

    Let us check if the proposed encoding scheme does achieve capacity.
    By summing $\rho(p)$ over all positions $p \in [\ell]$, we see that
    the total amount of data is the following multiplied by $n$:
    \begin{align}
        \sum_{p=1}^\ell \rho(p)
        & = \frac{\Cap(W^3)}{4} \cdot
            \Bigl(\frac{\log n}{\Cap(W^2)}
                - \frac{\log n}{\Cap(W^3)}\Bigr)
        \\ & \qquad + \frac{\Cap(W^3) + \Cap(W^2)}{4} \cdot
            \Bigl(\frac{\log n}{\Cap(W^1)}
                - \frac{\log n}{\Cap(W^2)}\Bigr)
        \\ & \qquad + \frac{\Cap(W^3) + \Cap(W^2) + \Cap(1)}{4} \cdot
            \Bigl(\ell
                - \frac{\log n}{\Cap(W^1)}\Bigr)
        \\ & = \frac{\Cap(W^3)}{4} \cdot
            \Bigl(\ell - \frac{\log n}{\Cap(W^3)}\Bigr)
        \\ & \qquad + \frac{\Cap(W^2)}{4} \cdot
            \Bigl(\ell - \frac{\log n}{\Cap(W^2)}\Bigr)
        \\ & \qquad + \frac{\Cap(W^1)}{4} \cdot
            \Bigl(\ell - \frac{\log n}{\Cap(W^1)}\Bigr)
        \\ & = \frac{\Cap(W^3) + \Cap(W^2) + \Cap(W^1)}{4} \cdot \ell
            - \frac34 \log n
    \end{align}
    This is nothing but expressing the area of an echelon shape by
    cutting it horizontally versus vertically.  The last line indeed
    matches the right-hand side of \eqref{eq:withP}, proving our claim.

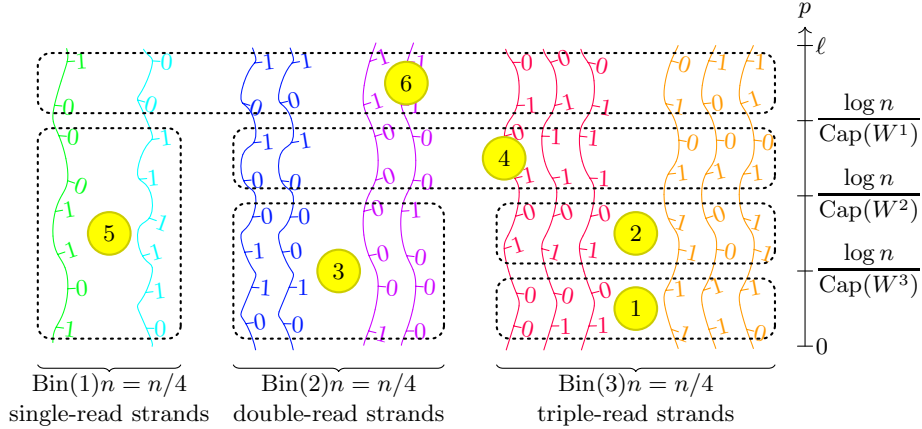
\begin{figure}
    \centering
    \footnotesize
    \begin{tikzpicture}[y=0.5cm]
        \pgfmathsetseed{8884040}
        \resetcolorseries[6]{no-y}
        \foreach \x in {1, ..., 6}{
            \tikzset{shift={({(\x+3)^2*0.125}, 0)}}
            \pgfmathsetmacro\phasex{rnd*360}
            \pgfmathsetmacro\frequx{random(50, 200)}
            \pgfmathsetmacro\phasey{rnd*360}
            \pgfmathsetmacro\frequy{\frequx*(1.6 + rnd*0.2)}
            \foreach \y in {1, ..., 8}{
                \foreach \xx in {1, ..., \numexpr\x/2} {
                    \tikzset{shift={(\xx*0.5, 0)}}
                    \ifnum \xx = 1
                        \pgfmathsetmacro\a{random(0, 1)}
                        \xdef\a{\a}
                    \fi
                    \pgfmathtruncatemacro\r{random(5)}
                    \ifnum \r = 1
                        \pgfmathsetmacro\a{random(0, 1)}
                    \fi
                    \draw [{no-y!![\x]}]
                        plot [domain=-0.5:0.5, samples=10, variable=\t]
                        ({wavex(\y+\t)}, {wavey(\y+\t)})
                        [shift={({wavex(\y)}, {wavey(\y)})}]
                        [rotate=rand*20, transform shape]
                        (0, 0) -- (0.1, 0)
                        node [right, inner sep=0] {$\a$}
                    ;
                }
            }
        }
        \tikzset{Step/.style={
            circle, fill=yellow, draw=yellow!80!black, solid
        }}
        \draw [dotted, line width=0.8pt, rounded corners=5pt]
            (12, 0.7) rectangle node [Step] {1} (8.3, 2.3)
            (12, 2.7) rectangle node [Step] {2} (8.3, 4.3)
            (7.6, 0.7) rectangle node [Step] {3} (4.8, 4.3)
            (12, 4.7) rectangle node [Step] {4} (4.8, 6.3)
            (4.1, 0.7) rectangle node [Step] {5} (2.2, 6.3)
            (12, 6.7) rectangle node [Step] {6} (2.2, 8.3)
        ;
        \draw [decorate, decoration=brace]
            (4.1, 0) -- node [below, align=center]
            {$\Bin(1)n = n/4$ \\ single-read strands}
            (2.2, 0)
        ;
        \draw [decorate, decoration=brace]
            (7.6, 0) -- node [below, align=center]
            {$\Bin(2)n = n/4$ \\ double-read strands}
            (4.8, 0)
        ;
        \draw [decorate, decoration=brace]
            (12, 0) -- node [below, align=center]
            {$\Bin(3)n = n/4$ \\ triple-read strands}
            (8.3, 0)
        ;
        \tikzset{shift={(12.4, 0)}}
        \draw [->] (0, 0.5) -- (0, 9) node [above] {$p$};
        \draw
            (0, 8.5) +(-0.1, 0) -- +(0.1, 0)
            node [right, inner sep=1pt] {$\ell$}
            (0, 6.5) +(-0.1, 0) -- +(0.1, 0)
            node [right, inner sep=1pt] {$\dfrac{\log n}{\Cap(W^1)}$}
            (0, 4.5) +(-0.1, 0) -- +(0.1, 0)
            node [right, inner sep=1pt] {$\dfrac{\log n}{\Cap(W^2)}$}
            (0, 2.5) +(-0.1, 0) -- +(0.1, 0)
            node [right, inner sep=1pt] {$\dfrac{\log n}{\Cap(W^3)}$}
            (0, 0.5) +(-0.1, 0) -- +(0.1, 0)
            node [right, inner sep=1pt] {$0$}
        ;
    \end{tikzpicture}
    \caption{
        The decoding part of Theorem~\ref{thm:useBR}.  Step 1: decode
        $\RRR$ to obtain indices.  Even number steps: subtract indices
        from $X$ and decode $\BBB$ to obtain data.  Odd number steps:
        subtract data from $X$ and decode $\RRR$ to obtain indices.
    }                                                 \label{fig:zigzag}
\end{figure}

\subsection{Decoding of the straw man}

    The decoding of our code follows an onion-peeling strategy that is
    depicted in Figure~\ref{fig:zigzag}.

    In step 1, we know that the first few positions does not contain any
    data.  So we look for triple-read strands and decode $\RRR(s)$ for
    their indices $s$ using the first few positions
    \[
        p \in \Bigl[
            1,
            \frac{\log n}{\Cap(W^3)}
        \Bigr].                                         \label{eq:step1}
    \]
    The decoding is bound to succeed because we have collected
    sufficiently many observations $Y^{S^{-1}(s)}_p$ that are generated
    by $W^3$.

    In step 2, since we have learned the indices of the triple-read
    strands, we can subtract the stream generated by $\RRR$ from those
    strands.  More formally, for every triple-read strand $s$, we
    compute
    \[
        L^s_p \coloneqq
        (-1)^{\RRR(s)_p} \sum_{r\in S^{-1}(s)} Y^r_p,
    \]
    and treat $L^s_p$ as the noisy observations of
    $\BBB^{\rho(p)}(\data^p)_s$ generated by $W^3$.  Here, we assume
    that $L^s_p$ and $Y^r_p$ are in the form of log-likelihood ratios
    (LLRs)\footnote{ We are using a property of LLRs that if we make
    multiple observations of the same bit, then the overall LLR is the
    sum of the LLRs.}.  Now for positions in this range
    \[
        p \in \Bigl[
            \frac{\log n}{\Cap(W^3)} + 1,
            \frac{\log n}{\Cap(W^2)}
        \Bigr],                                         \label{eq:step2}
    \]
    we can decode the $p$th chunk of data, $\data^p$, because the number
    of observations $L^s_p$ matches the code rate
    \[ \rho(p) \coloneqq \frac{\Cap(W^3)}{4} \]
    in this range.

    In step 3, we subtract the data chunks we just learned from the
    double-read strands.  That is,
    \[
        \Upsilon^s_p \coloneqq
        (-1)^{\BBB^{\rho(p)}(\data^p)_s}
        \sum_{r\in S^{-1}(s)} Y^r_p
    \]
    for range \eqref{eq:step2}.  The $\Upsilon$'s in range
    \eqref{eq:step2} and the $Y$'s in range \eqref{eq:step1} can be
    treated as noisy observations of $\RRR(s)_p$ generated by $W^2$.  We
    can decode the indices of the double-read strands as we have
    collected a sufficient number of observations.

    In step 4, we decode the data chunks in the range
    \[
        p \in \Bigl[
            \frac{\log n}{\Cap(W^3)} + 1,
            \frac{\log n}{\Cap(W^2)}
        \Bigr].                                         \label{eq:step4}
    \]
    We do so by subtracting the index stream $\RRR(s)_p$ from the
    double-read strands and the triple-read strands.  This yields noisy
    observations of $\BBB^{\rho(p)}(\data^p)$; $n/4$ of them are
    generated by $W^2$; the other $n/4$ by $W^3$.  Because the assigned
    code rate is
    \[ \rho(p) \coloneqq \frac{\Cap(W^3) + \Cap(W^2)}{4} \]
    in range \eqref{eq:step4}, we are guaranteed the recovery of the
    data chunks.

    Steps 5 and 6 follow the same zigzag pattern.  In step 5, we decode
    the indices of the single-read strands by first subtracting the data
    chunks we have learned in steps 2 and 4.  In steps 6, we decode the
    data chunks by first subtracting all indices we have learned in
    steps 1, 3, and 5.  Remark: Readers might argue that we actually do
    not need to encode the indices after $p > \log(n) / \Cap(W^1)$, and
    they would be right.

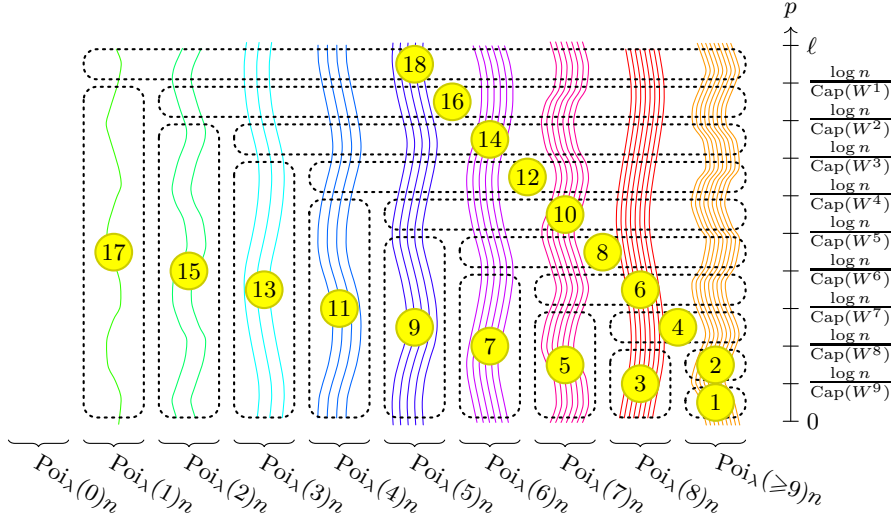
\begin{figure}
    \centering
    \footnotesize
    \begin{tikzpicture}[y=0.5cm]
        \pgfmathsetseed{8884848}
        \resetcolorseries[9]{no-y}
        \foreach \x in {1, ..., 9}{
            \tikzset{shift={(\x.5, 0)}}
            \pgfmathsetmacro\phasex{rnd*360}
            \pgfmathsetmacro\frequx{random(50, 200)}
            \pgfmathsetmacro\phasey{rnd*360}
            \pgfmathsetmacro\frequy{\frequx*(1.6 + rnd*0.2)}
            \path [save path=\wave]
                plot [domain=0:10, samples=20, smooth, variable=\t]
                ({wavex(\t)}, {wavey(\t)})
            ;
            \foreach \xx in {1, ..., \x} {
                \draw [{no-y!![\x]}, use path=\wave]
                    [transform canvas={
                        shift={(
                            {(\xx-0.5)/\x/2-1/4},
                            0
                        )}
                    }]
                ;
            }
        }
        \tikzset{Step/.style={
            circle, fill=yellow, draw=yellow!80!black, solid,
            inner sep=1pt, minimum size=1.5em
        }}
        \draw [dotted, line width=0.8pt, rounded corners=5pt]
            foreach \k in {1, 3, ..., 17} {
                (9.6-\k/2, \k/2+0.4) rectangle
                node [Step] {\k} (10.4-\k/2, 0.1)
            }
            foreach \k in {2, 4, ..., 18} {
                (10.1-\k/2, \k/2+0.9) rectangle
                node [Step] {\k} (9.9, \k/2+0.1)
            }
        ;
        \foreach  \k in {0, ..., 8} {
            \draw [decorate, decoration=brace]
                (\k.9, -0.5) -- node [below right, rotate=-30]
                {$\Poi_\lambda(\k)n$} (\k.1, -0.5)
            ;
        }
        \draw [decorate, decoration=brace]
            (9.9, -0.5) -- node [below right, rotate=-20]
            {$\Poi_\lambda({\geq}9)n$} (9.1, -0.5)
        ;
        \tikzset{shift={(10.5, 0)}}
        \draw [->] (0, 0) -- (0, 10.5) node [above] {$p$};
        \draw
            (0, 10) +(-0.1, 0) -- +(0.1, 0) node [right] {$\ell$}
            foreach \k in {1, ..., 9} {
                (0, 10-\k) +(-0.1, 0) -- +(0.1, 0)
                node [right] {$\frac{\log n}{\Cap(W^{\k})}$}
            }
            (0, 0) +(-0.1, 0) -- +(0.1, 0) node [right] {$0$}
        ;
    \end{tikzpicture}
    \caption{
        Step 1: decode indices.  Even number steps: subtract indices to
        decode data.  Odd number steps: subtract data to decode indices.
        This figure caps the number of reads at $\kappa \coloneqq 9$.
    }                                                  \label{fig:kappa}
\end{figure}

\subsection{Proof without assumptions}

    $\RRR$ and $\BBB^\rho$ are too good to be true in the assumptions we
    made earlier this section.  To prove Theorem~\ref{thm:useBR}
    unconditionally, we need to add deltas, epsilons, and w.h.p.s back.
    That is, a realistic block code $\BBB^\rho$ can recover the data
    from $(1 + \delta) \rho n$ observations with error probability
    \[
        \approx \exp(-\Omega(\delta^2 n)).             \label{eq:Berror}
    \]
    Meanwhile, a realistic rateless code $\RRR$ can recover a seed
    $s \in [n]$ from $(1 + \varepsilon) \* \log(n)/\Cap(W)$ observations
    generated by $W$ with error probabilities
    \[
        \approx \exp(-\Omega(\varepsilon^2 \log n))
        = n^{-\Omega(\varepsilon^2)}.                  \label{eq:Rerror}
    \]
    With these error probability bounds, we prove
    Theorem~\ref{thm:useBR} below.  The high-level structure of the
    proof will still piggyback the design of Figure~\ref{fig:zigzag}.
    See Figure~\ref{fig:kappa} for a more general depiction.

    We proceed to clarify how we take limits in the proof.  We will fix
    $\delta$ and $\varepsilon$ to be some small numbers while letting
    $\ell$, $m$, and $n$ goes to infinity.  The rate they go to infinity
    is such that $\ell/\log n$ and $\lambda \coloneqq m/n$ remain
    constant.  Afterwards, we will let $\delta$ and $\varepsilon$ go to
    zero while fixing a large number $\kappa$.  We use $\kappa$ to cap
    the number of reads each strand has; we will make the number of
    $kappa$-read strands $\approx \Poi_\lambda(\kappa+)$ by discarding
    any extra reads.  This $\kappa$ is the last constant go to infinity.

    Next, we move on to the code rate assignment.  For any $j \in
    [\kappa]$, we assign code rates
    \[
        \rho(p) \coloneqq
        \sum_{k=j}^\kappa \Poi_\lambda(k) \Cap(W^k) - 2\delta
        \quad\text{ for }\quad
        p \in \Bigl[
            \frac{(1 + \varepsilon) \log n}{\Cap(W^{j})},
            \frac{(1 + \varepsilon) \log n}{\Cap(W^{j-1})}
        \Bigr].
    \]
    Two edge cases are when $j = 1$ and $j = \kappa$.  For the former,
    $W^{j-1} = W^0$ is the garbage channel and the denominator
    $\Cap(W^{j-1})$ is zero; so the right end of the range should be
    replaced by the strand length $\ell$.  For the latter, $j = \kappa$,
    we enlarge the left end of the range to $0$.

    On the one hand, the $2\delta$ term in the assignment of $\rho(p)$
    will grant us the flexibility that we can still decode $\data^p$
    even if the number of $k$-read strands is not exactly
    $\Poi_\lambda(k) n$, but slightly fewer.  In fact, the probability
    that it deviates from $\Poi_\lambda(k) n$ by $\delta n$ is
    $\exp(-\Theta(\delta n))$, by a standard concentration argument.
    With such a small probability, and the fact that \eqref{eq:Berror}
    also decays exponentially in $n$, we can afford to use a union bound
    over all positions $p \in [\ell]$ and all strands $s \in [n]$.

    On the other hand, the $(1 + \varepsilon)$ term in the range of $p$
    will grant us the flexibility that we can decode the indices even if
    the rateless code is not so perfect.  This time, the error
    probability \eqref{eq:Rerror}, $n^{-\Omega(\delta^2)}$, decays
    polynomially in $n$---union bounds do not apply.

    To fix this, we actually do not need to change anything: Any
    $k$-read strand whose index cannot be decoded correctly after $(1 +
    \varepsilon) \log(n)/\Cap(W^k)$ observations will just be recognized
    as an observation of some random strand.  They will end up
    contributing misleading observations.  Let's just say that they are
    so misleading that they completely overwrite any other observation.
    But by using gap to capacity $2\delta$, the code can still tolerate
    another $\delta n$ errors before the recovery of $\data^p$ fail.  As
    a result of this, the probability that incorrect indices will cause
    any trouble is the probability that $n$ Bernoulli trials, each with
    success probability $n^{-\Omega(\delta^2)}$, succeeds at least
    $\delta n$ times.  Such a probability decays exponentially in $n$.

    So far we have elaborated on the error probabilities and argued that
    our scheme works with or without the unrealistic assumptions.  To
    compute the total amount of data, use a computation similar to
    Section~\ref{sec:strawrate},
    \begin{align}   
        & \kern-2em \sum_{j=1}^\kappa
            \Bigl(
                - 2\delta
                + \sum_{k=j}^\kappa \Poi_\lambda(k) \Cap(W^k)
            \Bigr)
            \Bigl( \text{length of the $j$th range} \Bigr)
        \\ & = - 2 \delta \ell
            + \sum_{k=1}^\kappa \Poi_\lambda(k) \Cap(W^k)
            \Bigl(
                \ell - \frac{(1 + \varepsilon) \log n} {\Cap(W^k)}
            \Bigr)
        \\ & \geq - 2 \delta \ell
            + \ell (\Cap(W^{\fish\lambda}) - \Poi_\lambda({>}\kappa))
        \\ &\kern2em - (1 + \varepsilon)
            (1 - \Poi_\lambda(0) - \Poi_\lambda({>}\kappa)) \log n.
    \end{align}
    As $\delta$ and $\varepsilon$ go to zero, the total amount of data
    simplifies to
    \[
        n \ell (\Cap(W^{\fish\lambda}) - \Poi_\lambda({>}\kappa))
        - (1 - \Poi_\lambda(0) - \Poi_\lambda({>}\kappa)) n \log n.
    \]
    As $\kappa$ goes to infinity, $\Poi_\lambda({>}\kappa)$ vanishes.
    And now it matches the capacity bound \eqref{eq:withP}.

    Leaving the search of $\RRR$ and $\BBB^\rho$ for future works, we
    hereby conclude the proof of Theorem~\ref{thm:useBR}.

\section{Future Works}                                \label{sec:future}

    In this paper, summarized by Figure~\ref{fig:useBR}, we weave
    rateless codes and block codes together to achieve the capacity of
    DNA coding.  Throughout the paper, we have made several black-box
    uses of good codes without specifying what they are.  A natural
    question is to find out what codes fit.  In particular, we are
    interested in the least possible encoding and decoding complexity
    that can achieve DNA's capacity.  We are also looking forwards to
    codes that can handle asymmetric channels and larger (preferably
    quaternary) alphabets.

    Another interesting direction that will strengthen the results is to
    decode without the genie revealing the partition $P$.  It appears to
    us that a strand length of $\ell > \log(n) / \Cap(W)$ is enough for
    an LLR-type distance to be able to cluster the reads with
    exponentially few errors.  We anticipate a formal treatment of this.

    The last elephant in the room is that DNA coding almost always comes
    with synchronization errors.  We have not addressed this issue in
    this paper, but we would suggest that a good inner code---a code
    that wraps up a sub-strand of $l \ll \ell$ letters as a $q^l$-ary
    symbol for outer codes to work on---should be able to handle
    synchronization errors as well as asymmetric errors.

\bibliographystyle{alpha}
\bibliography{GenoWeave-30}

\end{document}